\documentclass[11pt]{article}

\usepackage{a4}
\usepackage{pst-all}
\catcode`\@=11%
\usepackage{graphicx}
\usepackage{subfigure}
\usepackage{natbib}

\psset{unit=9.37mm,arrowscale=1.5}
\SpecialCoor
\usepackage{amssymb}
\usepackage{color}
\usepackage{amsmath}
\usepackage{url}

\usepackage{afterpage}
\usepackage{booktabs}
\usepackage{multirow}

\newcommand{\BPi}{\mbox{{\boldmath  $\Pi$}}}
\newcommand{\Bsigma}{\mbox{{\boldmath $\Sigma$}}}

\newcommand{\Bmu}{\mbox{{\boldmath  $\mu$}}}
\newcommand{\Bnu}{\mbox{{\boldmath  $\nu$}}}
\newcommand{\Bbeta}{\mbox{{\boldmath  $\beta$}}}

\newcommand{\transp}{{\sf T}}

\begin{document}

\title{Two Bayesian Approaches to \\ Dynamic Gaussian Bayesian Networks \\with Intra- and Inter-Slice Edges}

\date{}
\maketitle
\vspace{-1.5cm}

{\large\centering \textbf{Kezhuo Li and Marco Grzegorczyk}
	\vspace{+0.25cm} \normalsize
    
Bernoulli Institute (BI) \\
Rijksuniversiteit Groningen, NL \\
\emph{E-mail: m.a.grzegorczyk@rug.nl}  \\

\vspace{+0.5cm}}

\begin{abstract}
Gaussian Dynamic Bayesian Networks (GDBNs) are a widely used tool for learning network structures from continuous time-series data. To capture both time-lagged and contemporaneous dependencies, advanced GDBNs allow for dynamic inter-slice edges as well as static intra-slice edges. In the literature, two Bayesian modeling approaches have been developed for GDBNs. Both build on and extend the well-known Gaussian BGe score. We refer to them as the mean-adjusted BGe (mBGe) and the extended BGe (eBGe) models. In this paper, we contrast the two models and compare their performance empirically. The main finding of our study is that the two models induce different equivalence classes of network structures. In particular, the equivalence classes implied by the eBGe model are non-standard, and we propose a new variant of the DAG-to-CPDAG algorithm to identify them. To the best of our knowledge, these non-standard equivalence classes have not been previously reported.
\end{abstract}

\noindent {\bf Keywords} \\
Dynamic Bayesian networks; Gaussian Bayesian networks; equivalence classes; CPDAG; mBGe;  eBGe

\section{Introduction} 
\label{sec:introduction}
Dynamic Bayesian Networks (DBNs) are a widely used class of models for learning network structures from time-series data. Originally introduced by \cite{DBN_1}, DBNs have since been substantially developed, with seminal contributions by \cite{Fried_Murphy_DBN} and Murphy \cite{DBN_2,MURPHY_PHD}. A recent overview of DBNs and their various applications is provided by \cite{SCUTARI_NEERLANDICA}. To capture both time-lagged and contemporaneous dependencies, DBNs typically allow for dynamic inter-slice edges as well as static intra-slice edges. \\

\noindent In this paper, we focus on Gaussian DBNs (GDBNs), that is, DBNs for continuous time-series data. Two Bayesian modeling approaches for GDBNs have been developed in the literature. Both models build on and extend the well-known Gaussian BGe score of Geiger and Heckerman \cite{GeigerHeckGaussUAI,HECKERMAN_ANNALS,HECKERMAN_ADDENDUM}. We refer to these approaches as the  extended BGe (eBGe) model and the mean-adjusted BGe (mBGe) model. The eBGe model is implemented, for example, in the widely used {\bf R} packages {\em bnlearn} \cite{SCUTARI_BOOK} and {\em BiDAG} \cite{SOFTWARE3}. The mBGe model has more often been applied in the context of seemingly unrelated regression models \citep{SALAM_CS} and has recently been adapted for learning Bayesian networks from hybrid data \citep{Gryze2025}. Both the mBGe and eBGe models assume that the data arise from a multivariate Gaussian distribution, but they differ conceptually in how intra- and inter-slice dependencies are encoded. The eBGe model handles intra- and inter-slice dependencies jointly by augmenting the multivariate Gaussian distribution, so that network nodes and their lagged counterparts are represented together in a single enlarged Gaussian model. In contrast, the mBGe model encodes static (intra-slice) dependencies through the covariance structure and incorporates dynamic (inter-slice) dependencies via the mean vector. Detailed mathematical descriptions of the mBGe and eBGe models are provided in Sections~\ref{sec:mbge} and \ref{sec:ebge}, respectively.\\

\noindent In both GDBN models, all edges are directed. The dynamic inter-slice edges define the dynamic graph, whereas the static intra-slice dependencies define the static graph. While the dynamic graph is subject to no structural restrictions, the static graph is required to be a directed acyclic graph (DAG). DAGs fall into graph equivalence classes, a well-studied concept in the context of static Bayesian networks; see, for example, the works by Chickering \cite{Chickering_UAI95,CHICK_2002}. However, graph equivalence classes for DBNs with both inter- and intra-slice dependencies have received little attention. In particular, it appears not to have been recognized that the mBGe and eBGe models induce different graph equivalence classes. For the mBGe model, the correct equivalence class can be obtained in a very intuitive way by simply extracting the CPDAG of the static DAG. Since we could not find any literature source suggesting otherwise, we assume that most researchers extract graph equivalence classes under the more widely applied eBGe model in the same way. This is conceptually incorrect, since a non-standard DAG-to-CPDAG algorithm is required. However, commonly used Bayesian network software packages such as {\em bnlearn} \cite{SCUTARI_BNLEARN,SCUTARI_BNLEARN2,SCUTARI_BOOK} and {\em BiDAG} \cite{SOFTWARE3}, while supporting DBNs with both inter- and intra-slice edges, currently do not provide algorithms for correctly extracting CPDAGs under the eBGe model.\\

\noindent Under the eBGe model, static and dynamic dependencies are modeled simultaneously through an augmented Gaussian distribution. As a consequence, dynamic edges break the standard equivalence classes associated with the static DAG. We demonstrate this effect and show how standard DAG-to-CPDAG algorithms must be adapted in order to identify the correct ones under the eBGe model. The modified DAG-to-CPDAG algorithm proposed here is not only relevant for Gaussian DBNs, but is also required for extracting CPDAGs of discrete DBNs with both inter- and intra-slice edges. \\

\noindent This paper is organized as follows. After a brief review of Bayesian networks in Section~\ref{sec:cpdag}, with a particular focus on graph equivalence classes, we turn to dynamic Bayesian networks (DBNs) in Section~\ref{sec:methods}. Section~\ref{sec:BGE} reviews the traditional BGe score \citep{GeigerHeckGaussUAI,HECKERMAN_ANNALS,HECKERMAN_ADDENDUM}, after which we show in Section~\ref{sec:dgbn} how it is extended to the mBGe and eBGe models. In Section~\ref{sec:diff}, we compare the graph equivalence classes induced by the eBGe and mBGe models and propose an algorithm for extracting the CPDAG under the eBGe model. The data used for the method comparison and additional technical details are described in Sections~\ref{sec:data} and \ref{sec:technical}, respectively. The empirical results are presented in Section~\ref{sec:results}, and we conclude with a brief discussion in Section~\ref{sec:conclusions}.

 \section{Bayesian network equivalence classes}
\label{sec:cpdag}
Bayesian networks (BNs) use directed acyclic graphs (DAGs) to model the conditional independencies among random variables $X_1,\ldots,X_n$. Each variable $X_i$ is identified with a node of the DAG, and the set of directed edges among the $n$ nodes encodes the conditional independence relations among the variables. A node $X_j$ is called a parent of $X_i$ if there is a directed edge $X_j \rightarrow X_i$ pointing from $X_j$ to $X_i$. Given any DAG $\mathcal{G}$, let $\BPi_i$ denote the set of all parent nodes of $X_i$ induced by $\mathcal{G}$. Then $X_j$ is a parent of $X_i$ in $\mathcal{G}$ if and only if $X_j \in \BPi_i$. Moreover, we call $X_j$ an ancestor of $X_i$ if there exists a sequence of directed edges $X_j \rightarrow \cdots \rightarrow X_i$ leading from $X_j$ to $X_i$. Due to the acyclicity constraint, there are no cycles, that is, no sequences of the form $X_i \rightarrow \cdots \rightarrow X_i$.
In BNs, DAGs are used to encode conditional independencies such that the joint distribution factorizes into a product of local conditional distributions. More precisely, a DAG $\mathcal{G}$ with parent sets $\BPi_1,\ldots,\BPi_n$ implies
\begin{equation}
\label{eq:factorization}
p(X_1,\ldots,X_n | \mathcal{G}) = \prod_{i=1}^n p(X_i | \BPi_i).
\end{equation}
Although different DAGs may induce different parent sets, they can encode the same set of conditional independence relations among the variables. Such DAGs are said to be (Markov) equivalent, as they represent the same independence model. 
This notion induces DAG equivalence classes, such that all DAGs within a given equivalence class encode the same conditional independence relations among the variables \citep{Chickering_UAI95}. \cite{CHICK_2002} shows that two DAGs are equivalent if and only if they have (i) the same skeleton and (ii) the same set of v-structures:
\begin{itemize}
\item The skeleton of a DAG $\mathcal{G}$ is obtained by replacing all directed edges in $\mathcal{G}$ with undirected edges.
\item A v-structure is a configuration of the form $X_j \rightarrow X_i \leftarrow X_k$, where there is no edge between the parent nodes $X_j$ and $X_k$.
\end{itemize}
\cite{CHICK_2002} also shows that  DAG equivalence classes can be represented by `completed partially directed acyclic graphs' (CPDAGs). 
The DAGs within an equivalence class and the corresponding CPDAG share the same skeleton. However, unlike DAGs, a CPDAG may contain undirected edges.
\begin{itemize}
\item A directed edge $X_j \rightarrow X_i$ in a CPDAG is called a \emph{compelled} edge, since \emph{all} DAGs within the equivalence class agree on its orientation.
\item An undirected edge $X_j - X_i$ in a CPDAG is called a \emph{reversible} edge, since the DAGs within the equivalence class disagree on its orientation. All DAGs contain an edge between $X_i$ and $X_j$, but some have the edge $X_j \rightarrow X_i$, while others have the oppositely oriented edge $X_i \rightarrow X_j$.
\end{itemize}
\begin{figure}[t]\centering
\includegraphics[width=0.99\linewidth]{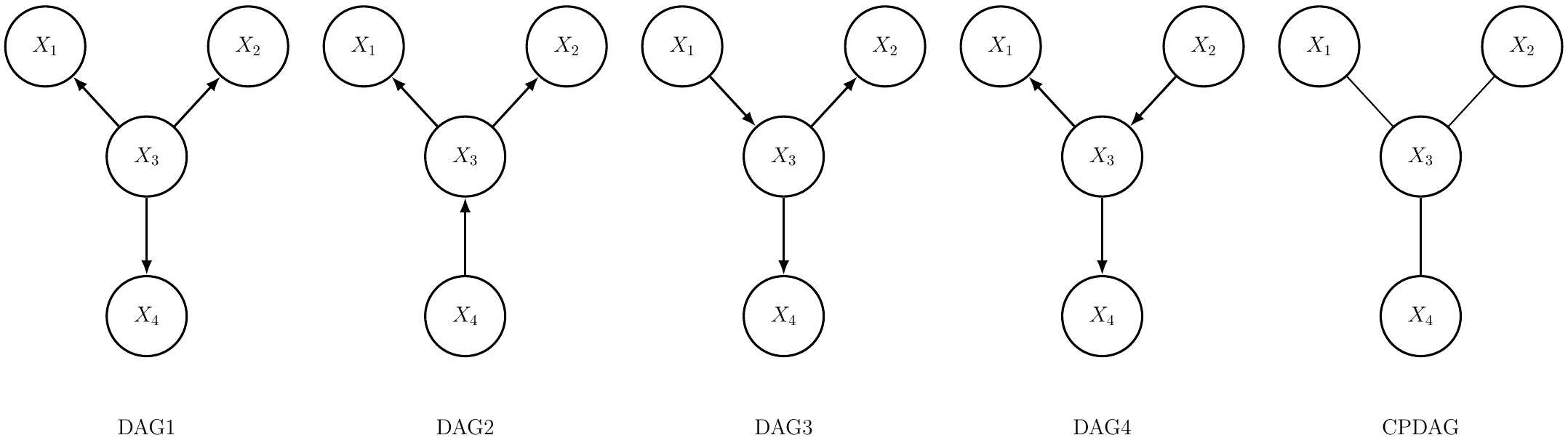}
\vspace{-0.3cm}
\caption{\footnotesize \textbf{Example:} Equivalent DAGs and their CPDAG. All three edges are reversible.}
\label{FIG1}
\end{figure}

\begin{figure}[t]\centering
\includegraphics[width=0.99\linewidth]{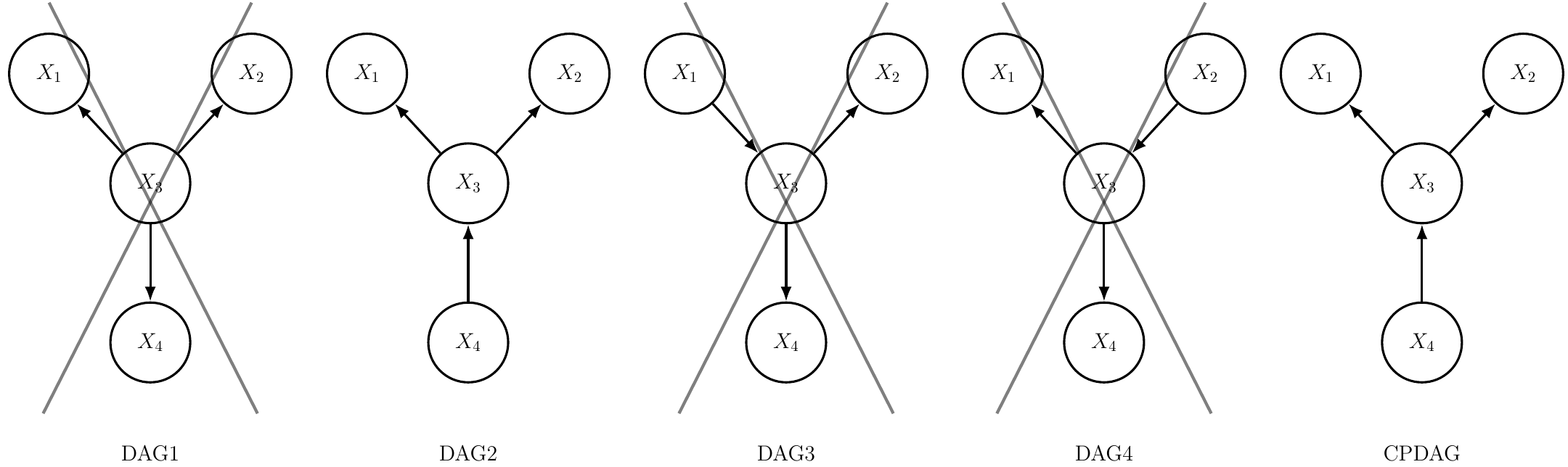}
\vspace{-0.3cm}
\caption{\footnotesize \textbf{Example from Figure~\ref{FIG1} under a structural constraint.} Restricting the edge between $X_4$ and $X_3$ to point towards $X_3$ renders all three edges compelled.}
\label{FIG2}
\end{figure}

With reference to Section~\ref{sec:diff}, consider a DAG containing the edge $X_j \rightarrow X_i$. The DAG belongs to an equivalence class, and the edge $X_j \rightarrow X_i$ is
\begin{itemize}
\item \emph{compelled} if and only if there exists no DAG in the equivalence class that contains the oppositely oriented edge $X_j \leftarrow X_i$;
\item \emph{reversible} if and only if there exists at least one DAG in the equivalence class that contains the oppositely oriented edge $X_j \leftarrow X_i$.
\end{itemize}
In Section~\ref{sec:diff}, we consider graphs that are subject to structural constraints, such that some edges are restricted to a single admissible direction. These constraints enforce certain edges to be compelled, as DAGs containing the oppositely oriented edges are invalid and therefore must be excluded from the equivalence class. As a result, structural constraints reduce the size of equivalence classes and may induce additional compelled edges.\\
An illustration is provided in Figures~\ref{FIG1}--\ref{FIG2}. Figure~\ref{FIG1} shows four equivalent DAGs and their corresponding CPDAG. The four DAGs share the same skeleton and contain no v-structures. Since each edge appears in both possible orientations across the DAGs, all three edges are reversible. Imposing the structural constraint that the edge between $X_4$ and $X_3$ may only be oriented towards $X_3$ renders three of the four DAGs invalid. Consequently, only a single DAG remains in the equivalence class, resulting in three compelled edges, as illustrated in Figure~\ref{FIG2}.

\section{Dynamic Gaussian Bayesian networks}
\label{sec:methods}
Gaussian Dynamic Bayesian Networks (GDBNs) can be used to model dependencies among continuous variables $X_1,\ldots,X_n$ observed at equidistant time points $t = 1,\ldots,T$. We consider GDBNs with both inter-slice (temporally lagged) and intra-slice (contemporaneous) edges among the variables.
In the literature, two conceptually distinct Bayesian approaches to GDBNs have been proposed. Both build on the traditional BGe scoring metric for intra-slice edges in static networks to accommodate inter-slice edges across time. In Section~\ref{sec:BGE}, we briefly review the traditional BGe score for static Gaussian Bayesian networks, before turning to the two extensions for GDBNs in Section~\ref{sec:dgbn}. The mean-adjusted BGe score (mBGe) and the extended BGe score (eBGe) for GDBNs are presented in Sections~\ref{sec:mbge} and~\ref{sec:ebge}, respectively. In Section~\ref{sec:diff}, we compare the two GDBN scores (mBGe vs.\ eBGe). To the best of our knowledge, these two approaches have not previously been contrasted, and an important difference concerning the induced graph equivalence classes has not been identified or reported in the literature.

\subsection{The traditional BGe score}
\label{sec:BGE}
In static Bayesian networks, directed acyclic graphs (DAGs) are used to encode conditional independence relations among the variables $X_1,\ldots,X_n$. We have $N$ independent observations, indexed by $i = 1,\ldots,N$, which are not temporally ordered. Any DAG $\mathcal{G}$ over the nodes $X_1,\ldots,X_n$ induces a parent set $\BPi_i$ for each variable $X_i$, and the joint distribution factorizes into a product of local conditional distributions:
\[
p(X_1,\ldots,X_n | \mathcal{G}) = \prod_{i=1}^n p(X_i | \BPi_i).
\]

In Gaussian static Bayesian networks (GSBNs), it is assumed that the random vector
${\bf X} = (X_1,\ldots,X_n)^{\transp}$ follows a multivariate Gaussian distribution.
Let ${\bf x}_i = (x_{i,1},\ldots,x_{i,n})^{\transp}$ denote the $i$th observed realization
of ${\bf X}$. The data form a random sample of size $N$, that is,
\begin{equation}
\label{EQ_VECTOR}
{\bf x}_i | (\Bmu,\Bsigma) \sim \mathcal{N}_n(\Bmu,\Bsigma),
\qquad (i = 1,\ldots,N).
\end{equation}

The BGe score \citep{HECKERMAN_ANNALS,HECKERMAN_ADDENDUM,GeigerHeckGaussUAI2,GeigerHeckGaussUAI} imposes a fully conjugate Normal--Wishart prior on the mean vector $\Bmu$ and the precision matrix ${\bf W} = \Bsigma^{-1}$:
\begin{eqnarray}
\label{BGE_prior_mix_old}
{\bf W} &\sim& \mathcal{W}_n(\alpha_{w},{\bf R}), \\
\nonumber
\Bmu |{\bf W} &\sim& \mathcal{N}_n\!\left(\Bnu,(\alpha_{\mu} {\bf W})^{-1}\right),
\end{eqnarray}
where the hyperparameters ${\bf R}\in\mathbb{R}^{n\times n}$ are positive definite,
$\Bnu \in \mathbb{R}^n$, $\alpha_{w} > n-1$, and $\alpha_{\mu} > 0$.
This fully conjugate prior yields the marginal likelihood
\begin{equation}
\label{MARGINAL_1}
p({\bf x}_{1:N})  =
\pi^{-\frac{nN}{2}}
\left(  \frac{\alpha_{\mu}}{\alpha_{\mu}+N} \right)^{\frac{n}{2}}
\frac{\Gamma_n\!\left(\frac{\alpha_{w}+N}{2}\right)}
     {\Gamma_n\!\left(\frac{\alpha_{w}}{2}\right)}
\frac{ \det({\bf R})^{\alpha_{w}/2} }
     { \det({\bf R} + {\bf T})^{(\alpha_{w}+N)/2} },
\end{equation}
where ${\bf x}_{1:N} := \{{\bf x}_1,\ldots,{\bf x}_N\}$ denotes the data,
$\Gamma_n(\cdot)$ is the multivariate gamma function, and
\begin{equation}
\label{T_ddagger_old}
{\bf T} :=
\sum_{i=1}^{N} ({\bf x}_{i} - \bar{{\bf x}}) ({\bf x}_{i} - \bar{{\bf x}})^{\transp}
+ \frac{\alpha_{\mu} N}{\alpha_{\mu}+N}
(\Bnu - \bar{{\bf x}}) (\Bnu - \bar{{\bf x}})^{\transp},
\end{equation}
with $\bar{{\bf x}}$ denoting the sample mean of ${\bf x}_1,\ldots,{\bf x}_N$.
Equation~\eqref{MARGINAL_1} corresponds to the marginal likelihood of the complete DAG,
in which all nodes are pairwise connected. For any $l$-dimensional subset of variables $L \subset \{X_1,\ldots,X_n\}$, the marginal likelihood can be computed as
\begin{eqnarray}
\label{MARGINAL_1_sub}
p({\bf x}_{1:N}^L)  =
\pi^{-\frac{lN}{2}}
\left( \frac{\alpha_{\mu}}{\alpha_{\mu}+N} \right)^{\frac{l}{2}}
\frac{\Gamma_l\!\left(\frac{\alpha_{w}-n+l+N}{2}\right)}
     {\Gamma_l\!\left(\frac{\alpha_{w}-n+l}{2}\right)}
\frac{ \det({\bf R}^{L,L})^{(\alpha_{w}-n+l)/2} }
     { \det({\bf R}^{L,L}+{\bf T}^{L,L})^{(\alpha_{w}-n+l+N)/2} },
\end{eqnarray}
where ${\bf x}_{1:N}^L$ denotes the data restricted to the $l$ variables in $L$, and
${\bf R}^{L,L}$ and ${\bf T}^{L,L}$ are the submatrices of ${\bf R}$ and ${\bf T}$ corresponding to the variables in $L$.
The marginal likelihood for any DAG $\mathcal{G}$ can then be computed using the factorization of Geiger and Heckerman \citep{GeigerHeckGaussUAI,HECKERMAN_ANNALS,HECKERMAN_ADDENDUM}:
\begin{equation}
\label{EQ_MARGINAL_ANY}
p({\bf x}_{1:N} | \mathcal{G})
= \prod_{i=1}^n
\frac{ p({\bf x}_{1:N}^{\{X_i,\BPi_i\}}) }
     { p({\bf x}_{1:N}^{\{\BPi_i\}}) },
\end{equation}
where $\BPi_i$ denotes the parent set of $X_i$ implied by $\mathcal{G}$. The marginal likelihoods for the variable subsets $\{X_i,\BPi_i\}$ and $\{\BPi_i\}$ are computed using Eq.~\eqref{MARGINAL_1_sub}.

\subsection{Extending the BGe score to GDBNs}
\label{sec:dgbn}

GDBNs use static and dynamic edges to represent intra-slice and inter-slice dependencies among the variables. As in Section~\ref{sec:BGE}, the static edges define a DAG $\mathcal{G}$ over the nodes, inducing a static parent set $\BPi_i^S$ for each variable $X_i$. Specifically, $X_j \in \BPi_i^S$ if and only if there is a static edge $X_{j,t} \rightarrow X_{i,t}$. In the same vein, let $\mathcal{G}^D$ denote the graph formed by the dynamic edges. Unlike $\mathcal{G}$, which must be acyclic, $\mathcal{G}^D$ is not subject to acyclicity constraints. The graph $\mathcal{G}^D$ induces a dynamic parent set $\BPi_i^D$ for each variable $X_i$. We have $X_j \in \BPi_i^D$ if and only if there is a dynamic edge $X_{j,t-1} \rightarrow X_{i,t}$. A GDBN observed at $T$ time points therefore yields $T-1$ effective observations, due to the one-step time lag.

\subsubsection{The mean-adjusted BGe score (mBGe)}
\label{sec:mbge}

In the mean-adjusted BGe score (mBGe), dynamic dependencies are incorporated through multivariate linear regression. Specifically, the mean vector $\Bmu$ in Eq.~\eqref{EQ_VECTOR} is assumed to depend on the values of $X_1,\ldots,X_n$ at the previous time point, and therefore varies over time. Formally,
\begin{equation}
\label{FIRST_EQ_MBGE}
{\bf x}_t | (\Bmu_t,\Bsigma) \sim \mathcal{N}_n(\Bmu_t,\Bsigma),
\qquad (t = 2,\ldots,T),
\end{equation}
where $\Bmu_t = \Bmu_t({\bf x}_{t-1})$, and $T$ denotes the number of equidistant time points. For notational simplicity, we suppress in the remainder of this paper the explicit dependence of $\Bmu_t$ on ${\bf x}_{t-1}$. Each component of the mean vector
\[
\Bmu_t = (\mu_{1,t},\ldots,\mu_{n,t})^{\transp}
\]
is a linear function of its dynamic parents. Specifically, the conditional expectation of $X_i$ at time $t$ is given by
\begin{equation}
\label{EQ:MEAN_t}
\mu_{i,t}
:= \beta_{i,0} + \sum_{j: X_j \in \BPi_i^D} x_{j,t-1}\,\beta_{i,j},
\end{equation}
where $\BPi_i^D$ denotes the dynamic parent set of $X_i$, and
\[
\Bbeta_i = \bigl(\beta_{i,0}, \{\beta_{i,j} : X_j \in \BPi_i^D\}\bigr)
\]
collects the corresponding regression coefficients. Here, $\beta_{i,0}$ is an intercept term, and each $\beta_{i,j}$ quantifies the contribution of the lagged variable $X_{j,t-1}$ to the conditional mean of $X_{i,t}$. Thus, $\Bbeta_i$ characterizes how the values of the dynamic parent nodes at time $t-1$ influence the expected value of $X_i$ at time $t$. For notational simplicity, we suppress the explicit dependence of $\Bbeta_i$ on $\BPi_i^D$. Equation~\eqref{EQ:MEAN_t} can be written more compactly as an inner product,
\begin{equation}
\label{EQ:MEAN_t2}
\mu_{i,t} = {\bf z}_{i,t-1}^{\transp}\,\Bbeta_i,
\end{equation}
where ${\bf z}_{i,t-1}$ is a vector starting with a leading $1$ (corresponding to the intercept term $\beta_{i,0}$) and followed by the relevant lagged values of ${\bf x}_{t-1}$, ordered so that each entry aligns with its associated regression coefficient in $\Bbeta_i$. Collecting all mean components yields the multivariate regression representation
\begin{equation}
\label{EQ:MEAN}
\Bmu_t := {\bf Z}_{t-1}\,\tilde{\Bbeta},
\end{equation}
where $\mathcal{G}^D = \{\BPi_1^D,\ldots,\BPi_n^D\}$ and
\begin{equation}
\nonumber
\Bmu_t :=
\begin{pmatrix}
\mu_{1,t} \\ \vdots \\ \mu_{n,t}
\end{pmatrix},
\qquad
{\bf Z}_{t-1} :=
\begin{pmatrix}
{\bf z}_{1,t-1}^{\transp} &        & {\bf 0} \\
                          & \ddots &         \\
{\bf 0}                   &        & {\bf z}_{n,t-1}^{\transp}
\end{pmatrix},
\qquad
\tilde{\Bbeta} :=
\begin{pmatrix}
\Bbeta_1 \\ \vdots \\ \Bbeta_n
\end{pmatrix}.
\end{equation}

This implies for $t = 2,\ldots,T$ that
\[
{\bf x}_t | (\mathcal{G}^D,\tilde{\Bbeta},\Bsigma)
\sim \mathcal{N}_n\!\left( {\bf Z}_{t-1}\tilde{\Bbeta},\,\Bsigma \right).
\]

Defining
\[
\operatorname{vec}({\bf x}_{2:T}) := ({\bf x}_2^{\transp},\ldots,{\bf x}_T^{\transp})^{\transp}
\]
yields the stacked regression model
\begin{equation}
\label{eq:stacked_regression}
\operatorname{vec}({\bf x}_{2:T}) | (\mathcal{G}^D,\tilde{\Bbeta},\Bsigma)
\sim \mathcal{N}_{(T-1)n}\!\left(
{\bf Z}\tilde{\Bbeta},
{\bf I} \otimes \Bsigma
\right),
\end{equation}
where ${\bf Z} := ({\bf Z}_{1}^{\transp},\ldots,{\bf Z}_{T-1}^{\transp})^{\transp}$.

A Gaussian prior is imposed on $\tilde{\Bbeta}$,
\begin{equation}
\label{eq_reg_prior}
\tilde{\Bbeta} | \lambda \sim \mathcal{N}_{\kappa}({\bf 0}, \lambda^{2} {\bf I}),
\end{equation}
where $\kappa = \sum_{i=1}^n (|\BPi_i^D|+1)$ is the length of $\tilde{\Bbeta}$ and $\lambda^2$ is a fixed hyperparameter. Integrating out $\tilde{\Bbeta}$ yields the marginal likelihood and full conditional distributions \cite{SALAM_CS,Gryze2025}:
\begin{eqnarray}
\label{EQ_MARGINAL_ALL}
\operatorname{vec}({\bf x}_{2:T}) | (\mathcal{G}^D,\Bsigma)
&\sim&
\mathcal{N}_{(T-1)n}\!\left(
{\bf 0},
{\bf I}\otimes\Bsigma + \lambda^2 {\bf Z}{\bf Z}^{\transp}
\right), \\
\label{FCD_1}
\tilde{\Bbeta} | (\mathcal{G}^D,\Bsigma,{\bf x}_{2:T})
&\sim&
\mathcal{N}_{\kappa}\!\left(
\Bsigma^{\star}{\bf Z}^{\transp}({\bf I}\otimes\Bsigma)^{-1}\operatorname{vec}({\bf x}_{2:T}),
\Bsigma^{\star}
\right),
\end{eqnarray}
where
\[
\Bsigma^{\star}
:= \left(\lambda^{-2}{\bf I}
+ {\bf Z}^{\transp}({\bf I}\otimes\Bsigma)^{-1}{\bf Z}
\right)^{-1}.
\]

Given $\mathcal{G}^D$ and $\tilde{\Bbeta}$, the mean vectors
\[
\Bmu_{2:T} := \{\Bmu_2,\ldots,\Bmu_T\}
\]
can be computed. Defining ${\bf y}_t := {\bf x}_t - \Bmu_t$, Eq.~\eqref{FIRST_EQ_MBGE} is equivalent to
\begin{equation}
\label{eq:y_definition}
{\bf y}_t | \Bsigma \sim \mathcal{N}_n({\bf 0},\Bsigma),
\qquad t = 2,\ldots,T.
\end{equation}

For this zero-mean Gaussian model, the simplified BGe score \cite{Gryze2024} can be applied. Equation~\eqref{MARGINAL_1_sub} then becomes
\begin{equation}
\label{EQ_MARGINAL2}
p({\bf y}_{2:T}^{L})
=
\pi^{-\frac{l(T-1)}{2}}
\frac{\Gamma_l\!\left(\frac{\alpha_w-n+l+(T-1)}{2}\right)}
     {\Gamma_l\!\left(\frac{\alpha_w-n+l}{2}\right)}
\frac{\det({\bf R}^{L,L})^{\frac{\alpha_w-n+l}{2}}}
     {\det({\bf R}^{L,L}+{\bf S}^{L,L})^{\frac{\alpha_w-n+l+(T-1)}{2}}},
\end{equation}
with
\[
{\bf S} := \sum_{t=2}^T ({\bf x}_t-\Bmu_t)({\bf x}_t-\Bmu_t)^{\transp}.
\]
For any DAG $\mathcal{G}$, this yields
\begin{equation}
\label{EQ_MARGINAL_ANY_NOVEL}
p({\bf y}_{2:T}|\mathcal{G})
=
\prod_{i=1}^n
\frac{p({\bf y}_{2:T}^{\{X_i,\BPi_i^S\}})}
     {p({\bf y}_{2:T}^{\{\BPi_i^S\}})},
\end{equation}
where $\BPi_i^S$ denotes the static parent set of $X_i$ induced by $\mathcal{G}$.

\paragraph{Markov chain Monte Carlo (MCMC) inference.}
Using an MCMC sampling scheme, the graphs $\mathcal{G}$ and $\mathcal{G}^D$, as well as the model parameters $\tilde{\Bbeta}$ and $\Bsigma$, can be sampled from the posterior distribution.
\begin{itemize}
\item[\textbf{1.}] Given $\mathcal{G}^D$ and $\tilde{\Bbeta}$, the simplified BGe score in Eq.~\eqref{EQ_MARGINAL_ANY_NOVEL} can be computed for any DAG $\mathcal{G}$, allowing $\mathcal{G}$ to be updated via single-edge operations \cite{MadiganYork,GiudiciMCMC}. For each DAG $\mathcal{G}$, a covariance matrix $\Bsigma$ consistent with $\mathcal{G}$ can be sampled using the algorithm of \cite{Gryze2023}.
\item[\textbf{2.}] Given the current $\Bsigma$, the marginal likelihood of each $\mathcal{G}^D$ can be computed using Eq.~\eqref{EQ_MARGINAL_ALL}. Single-edge operations are then used to update $\mathcal{G}^D$. Given the updated $\mathcal{G}^D$ and $\Bsigma$, the regression coefficients $\tilde{\Bbeta}$ are sampled from the full conditional distribution in Eq.~\eqref{FCD_1}.
\end{itemize}

\subsubsection{The extended BGe (eBGe) score}
\label{sec:ebge}

In the extended BGe score (eBGe), dynamic dependencies are incorporated through augmentation of the covariance structure.
Let $X_{i,t}$ denote variable $X_i$ at time $t$. Given a static DAG $\mathcal{G}$ and a dynamic graph $\mathcal{G}^D$, let $\BPi_{i,t}^S$ denote the static parents of $X_i$ at time $t$, and let $\BPi_{i,t-1}^D$ denote the dynamic parents of $X_i$ at time $t-1$. This implies
\begin{eqnarray}
\nonumber
\BPi_{i,t}^S &\subset& \{X_{1,t},\ldots,X_{n,t}\}, \\
\nonumber
\BPi_{i,t-1}^D &\subset& \{X_{1,t-1},\ldots,X_{n,t-1}\}.
\end{eqnarray}
The extended BGe (eBGe) score builds on the factorization
\[
p(X_{1,t},\ldots,X_{n,t}| X_{1,t-1},\ldots,X_{n,t-1},\mathcal{G},\mathcal{G}^D)
= \prod_{i=1}^n p(X_{i,t}| \BPi_{i,t}^S,\BPi_{i,t-1}^D).
\]

The eBGe score augments the variables $X_{1,t},\ldots,X_{n,t}$ with the lagged variables
$X_{1,t-1},\ldots,X_{n,t-1}$ and assumes that the resulting augmented vectors follow a joint
$2n$-dimensional Gaussian distribution. For the data, this implies
\begin{equation}
\label{eq:ebge}
\begin{pmatrix}
{\bf x}_{2} \\[1mm]
{\bf x}_{1}
\end{pmatrix},\ldots,
\begin{pmatrix}
{\bf x}_{T} \\[1mm]
{\bf x}_{T-1}
\end{pmatrix}
\sim
\mathcal{N}_{2n}\!\left(
\begin{pmatrix}
\Bmu_S \\[1mm]
\Bmu_D
\end{pmatrix},
\begin{pmatrix}
\Bsigma_{S,S} & \Bsigma_{S,D} \\
\Bsigma_{D,S} & \Bsigma_{D,D}
\end{pmatrix}
\right).
\end{equation}

Loosely speaking, the covariance block $\Bsigma_{S,S}$ must be consistent with the conditional
independence relations implied by the static DAG $\mathcal{G}$ and describes contemporaneous
intra-slice dependencies. The block $\Bsigma_{D,S}$ (with
$\Bsigma_{S,D}=\Bsigma_{D,S}^{\transp}$) must be consistent with the dynamic graph
$\mathcal{G}^D$ and captures temporally lagged inter-slice dependencies. Although model
stationarity would require the parameter constraints $\Bmu_S=\Bmu_D$ and
$\Bsigma_{S,S}=\Bsigma_{D,D}$, these constraints are ignored in the eBGe model.
For notational convenience, we relabel the variables and augmented
vectors as
\[
Z_{1,t}:=X_{1,t},\ldots,Z_{n,t}:=X_{n,t},\qquad
Z_{n+1,t}:=X_{1,t-1},\ldots,Z_{2n,t}:=X_{n,t-1},
\]
and define
\[
{\bf z}_t :=
(z_{1,t},\ldots,z_{n,t},z_{n+1,t},\ldots,z_{2n,t})^{\transp}
=
\begin{pmatrix}
{\bf x}_t \\[1mm]
{\bf x}_{t-1}
\end{pmatrix}
\in \mathbb{R}^{2n}.
\]

Since we assume time-homogeneous parent sets, we drop the time index and write for the parent sets in the augmented graph:
\begin{eqnarray}
\nonumber
\BPi_i^S &\subset& \{Z_1,\ldots,Z_n\}, \\
\nonumber
\BPi_i^D &\subset& \{Z_{n+1},\ldots,Z_{2n}\},
\end{eqnarray}
 and define the augmented data as ${\bf z}_{2:T}:=\{{\bf z}_2,\ldots,{\bf z}_T\}$. The eBGe model imposes a $2n$-dimensional Normal--Wishart prior
\begin{eqnarray}
\label{eq_aug_p1}
{\bf W}
:= 
\begin{pmatrix}
\Bsigma_{S,S} & \Bsigma_{S,D} \\
\Bsigma_{D,S} & \Bsigma_{D,D}
\end{pmatrix}^{-1}
&\sim&
\mathcal{W}_{2n}(\alpha_w,\tilde{\bf R}), \\
\label{eq_aug_p2}
\begin{pmatrix}
\Bmu_S \\[1mm]
\Bmu_D
\end{pmatrix}
| {\bf W}
&\sim&
\mathcal{N}_{2n}\!\left(\tilde{\Bnu},(\alpha_\mu{\bf W})^{-1}\right).
\end{eqnarray}
A limitation of the Normal--Wishart prior is that it does not enforce stationarity. For instance, if a node $X_j$ is both a static and a dynamic parent of $X_i$, then $X_{j,t}$ and $X_{j,t-1}$ appear jointly in the $i$th local conditional distribution. Since this prior does not impose equality of the means and variances of $X_{j,t}$ and $X_{j,t-1}$, the resulting local conditional distribution is miss-specified relative to stationarity assumptions and may lead to biased inference. One possible workaround is to disallow the simultaneous presence of both edges, so that $X_j$ cannot be a static and a dynamic parent of $X_i$ at the same time. Similar constraints arise in the presence of self-loops such as $X_{i,t-1}\rightarrow X_{i,t}$, which can also be ruled out.
\\
The eBGe marginal likelihood is given by:
\begin{equation}
\label{EQ_MARGINAL_ANY_NEW}
p({\bf z}_{2:T}|\mathcal{G},\mathcal{G}^D)
=
\prod_{i=1}^n
\frac{p({\bf z}_{2:T}^{\{Z_i,\BPi_i^S,\BPi_i^D\}})}
     {p({\bf z}_{2:T}^{\{\BPi_i^S,\BPi_i^D\}})},
\end{equation}
and for any $l$-dimensional variable subset $L\subset\{Z_1,\ldots,Z_{2n}\}$, we have:
{\scriptsize
\begin{eqnarray}
\nonumber
p({\bf z}_{2:T}^L)
&=&
\pi^{-\frac{l(T-1)}{2}}
\left(\frac{\alpha_\mu}{\alpha_\mu+(T-1)}\right)^{\frac{l}{2}}
\frac{\Gamma_l\!\left(\frac{\alpha_w-2n+l+(T-1)}{2}\right)}
     {\Gamma_l\!\left(\frac{\alpha_w-2n+l}{2}\right)}
\frac{\det(\tilde{\bf R}^{L,L})^{(\alpha_w-2n+l)/2}}
     {\det(\tilde{\bf R}^{L,L}+\tilde{{\bf T}}^{L,L})^{(\alpha_w-2n+l+(T-1))/2}},
\end{eqnarray}}
where
\begin{equation}
\nonumber
\tilde{{\bf T}}
:=
\sum_{t=2}^{T} ({\bf z}_t-\bar{\bf z})({\bf z}_t-\bar{\bf z})^{\transp}
+
\frac{\alpha_\mu(T-1)}{\alpha_\mu+(T-1)}
(\tilde{\Bnu}-\bar{\bf z})(\tilde{\Bnu}-\bar{\bf z})^{\transp},
\end{equation}
and $\bar{\bf z}$ denotes the sample mean of ${\bf z}_{2:T}$.\\
Because static and dynamic dependencies are modeled jointly through an augmented covariance structure, the eBGe score induces equivalence classes that differ from those implied by standard static DAG theory, as discussed in Section~\ref{sec:diff}.

\subsection{Graph equivalence classes (mBGe vs.\ eBGe)}
\label{sec:diff}
In both cases (mBGe and eBGe), an $n\times n$ covariance matrix $\Bsigma$ is used to describe
contemporaneous intra-slice dependencies ($X_{j,t}\rightarrow X_{i,t}$), and $\Bsigma$ must be
consistent with the static DAG $\mathcal{G}$. The two metrics differ in how they incorporate
dynamic inter-slice dependencies ($X_{j,t-1}\rightarrow X_{i,t}$):
\begin{itemize}
\item mBGe incorporates dynamic dependencies through the mean vector;
\item eBGe incorporates them by augmenting the covariance matrix.
\end{itemize}

All dynamic inter-slice edges ($X_{j,t-1}\rightarrow X_{i,t}$) are compelled in
their directions, since edges pointing backward in time would violate the assumed temporal ordering, according to which causes must precede their effects. By contrast, as described in Section~\ref{sec:cpdag}, static intra-slice edges ($X_{j,t}\rightarrow X_{i,t}$) may be either compelled or reversible.
The different treatment of dynamic edges in mBGe and eBGe  has a direct impact on the
 graph equivalence classes.

\begin{itemize}
\item \textbf{mBGe:}
Dynamic edges affect the model only through the mean vector, and static DAGs are inferred among
the mean-adjusted variables $Y_{1,t},\ldots,Y_{n,t}$, where
$Y_{i,t}:=X_{i,t}-\mu_{i,t}$ and $\mu_{i,t}$ captures the effect of all dynamic parents of $X_i$.
Since static and dynamic edges never jointly appear, the static DAG among
$Y_{1,t},\ldots,Y_{n,t}$ is subject to the standard equivalence classes
\cite{Gryze2024}. Consequently, CPDAGs can be computed using standard DAG-to-CPDAG algorithms
\cite{Chickering_UAI95,CHICK_2002}. The final CPDAG is obtained by merging the dynamic graph (with all dynamic edges fixed) with the static CPDAG.

\item \textbf{eBGe:}
Static and dynamic edges are treated symmetrically and appear jointly in the
augmented graph defined over the $2n$ nodes
$X_{1,t},\ldots,X_{n,t}$ and $X_{1,t-1},\ldots,X_{n,t-1}$. As a result, graph equivalence cannot be
characterised solely in terms of the static DAG among $X_{1,t},\ldots,X_{n,t}$.
In particular, computing the CPDAG of the static DAG alone is insufficient.
Moreover, computing the CPDAG of the augmented graph and subsequently extracting the sub-CPDAG
over $X_{1,t},\ldots,X_{n,t}$ is also incorrect, since all dynamic edges of the form
$X_{j,t-1}\rightarrow X_{i,t}$ are subject to structural constraints and must be compelled to
point forward in time. These constraints can render static edges compelled in the augmented graph
that would be reversible in the static DAG. The correct procedure is therefore
to extract the CPDAG of the augmented graph while explicitly enforcing all dynamic edges to be
compelled.
\end{itemize}

\begin{figure}[t]\centering \includegraphics[width=0.99\linewidth]{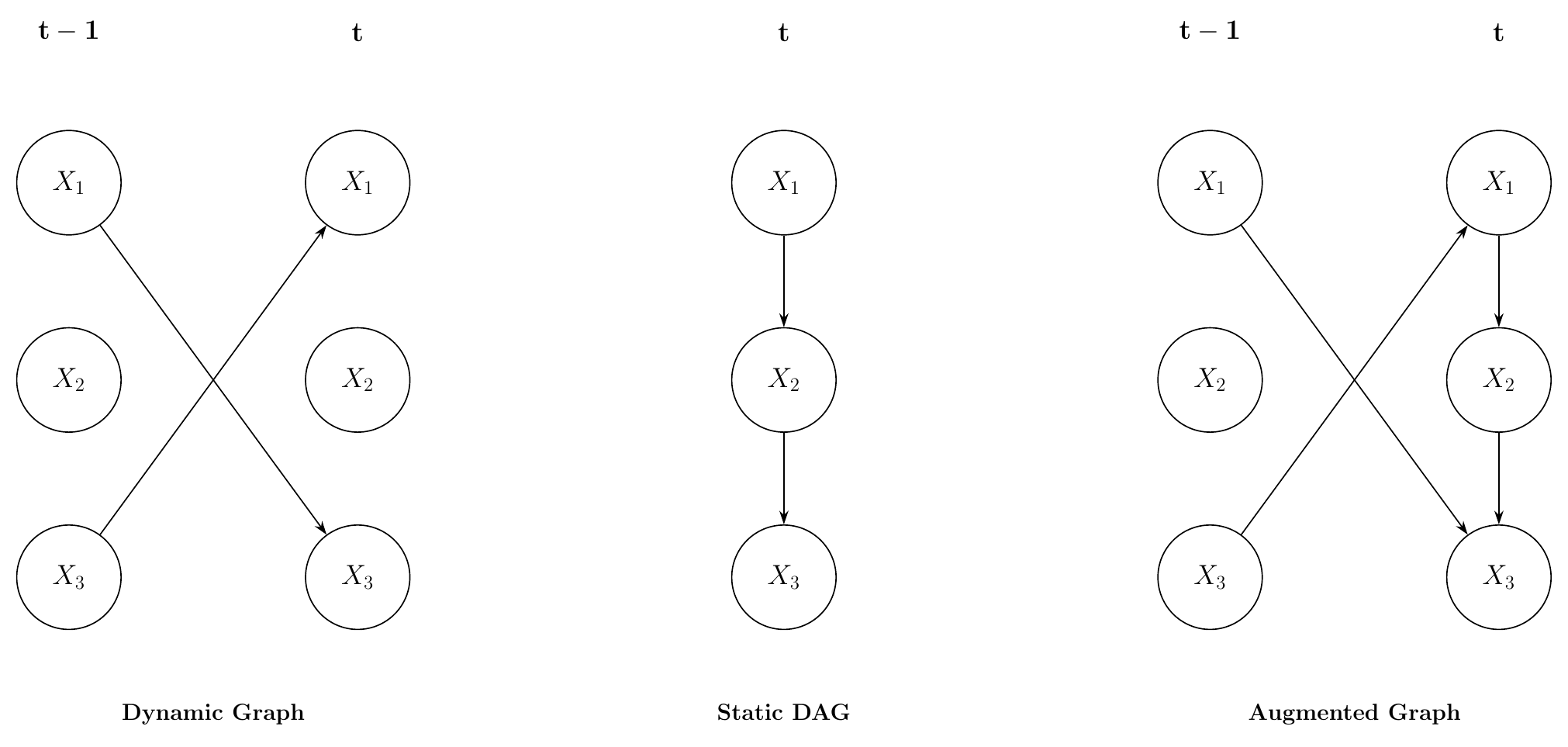} \vspace{-0.3cm} \caption{\footnotesize {\bf Graphical illustration (1 of 3).} GDBN with three nodes. The figure shows the dynamic graph (left), the static DAG (center), and the augmented graph (right).} \label{FIG3} \end{figure} 

\begin{figure}[t]\centering \includegraphics[width=0.99\linewidth]{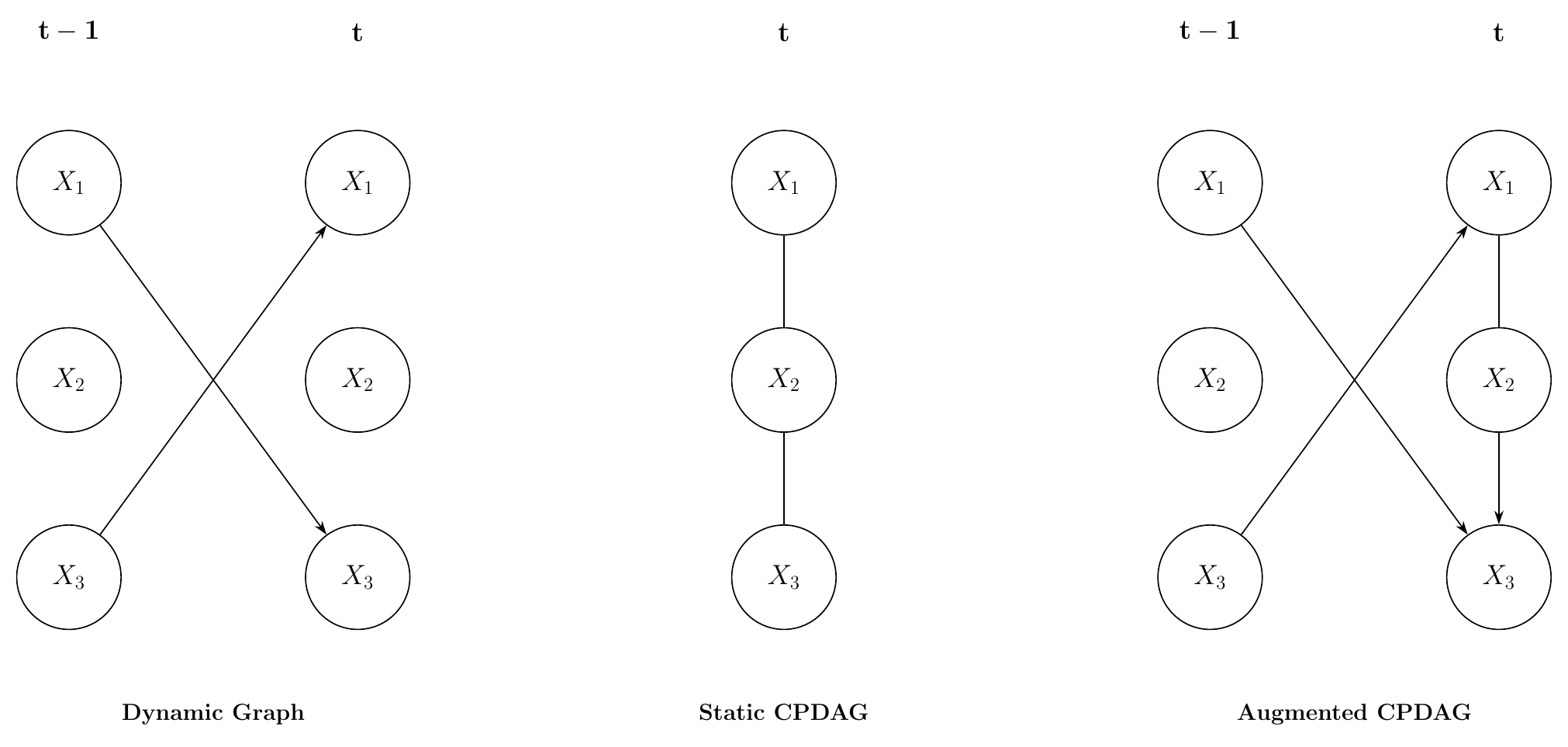} \vspace{-0.3cm} \caption{\footnotesize {\bf Graphical illustration (2 of 3).} The dynamic graph (left), the static CPDAG (center), and the CPDAG of the augmented graph (right) of the GDBN from Figure~\ref{FIG3}.} \label{FIG4} \end{figure} 

\begin{figure}[t]\centering \includegraphics[width=0.99\linewidth]{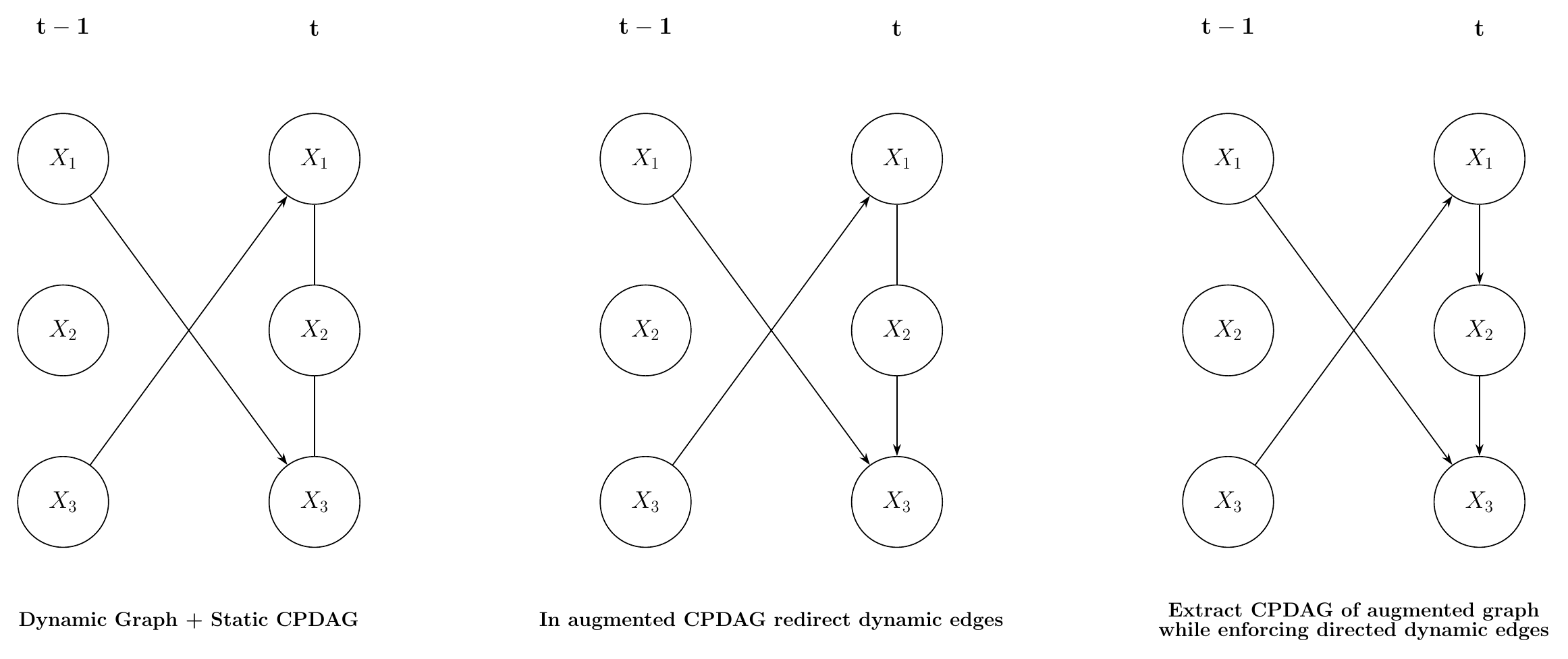} \vspace{-0.3cm} \caption{\footnotesize {\bf Graphical illustration (3 of 3).} The mBGe CPDAG (left), the eBGe CPDAG (right), and the naive augmented CPDAG (center). } \label{FIG5} \end{figure}

\noindent {\bf Illustrative Example:} Consider a small example GDBN with $n=3$ nodes $X_1$, $X_2$, and $X_3$. The dynamic graph and
the static DAG are shown in the left and center panels of Figure~\ref{FIG3}, while the
corresponding augmented graph used by the eBGe score is shown in the right panel. The CPDAG of
the static DAG (center panel) and the CPDAG of the augmented graph (right panel) are displayed
in Figure~\ref{FIG4}. For the dynamic graph (left panel), there are no equivalence classes, since
dynamic edges can only point forward in time.

\begin{itemize}
\item \textbf{mBGe:}
The correct CPDAG is obtained by combining the dynamic graph (with all dynamic edges fixed) with the static CPDAG, yielding the
graph shown in the left panel of Figure~\ref{FIG5}. 

\item \textbf{eBGe:}
The correct CPDAG is obtained by extracting the CPDAG of the augmented graph shown in the right
panel of Figure~\ref{FIG4}, while explicitly enforcing the two dynamic edges to point forward in
time (i.e., to be compelled). This yields the graph shown in the right panel of
Figure~\ref{FIG5}.

\item \textbf{Naive augmentation:}
The CPDAG of the augmented graph without enforcing the time-direction constraints is shown in
the right panel of Figure~\ref{FIG4}. Redirecting its dynamic edges post hoc yields the graph
shown in the center panel of Figure~\ref{FIG5}. This graph coincides with neither the mBGe CPDAG nor the eBGe CPDAG.
\end{itemize}

   \begin{figure}[p]\centering
\includegraphics[width=0.99\linewidth]{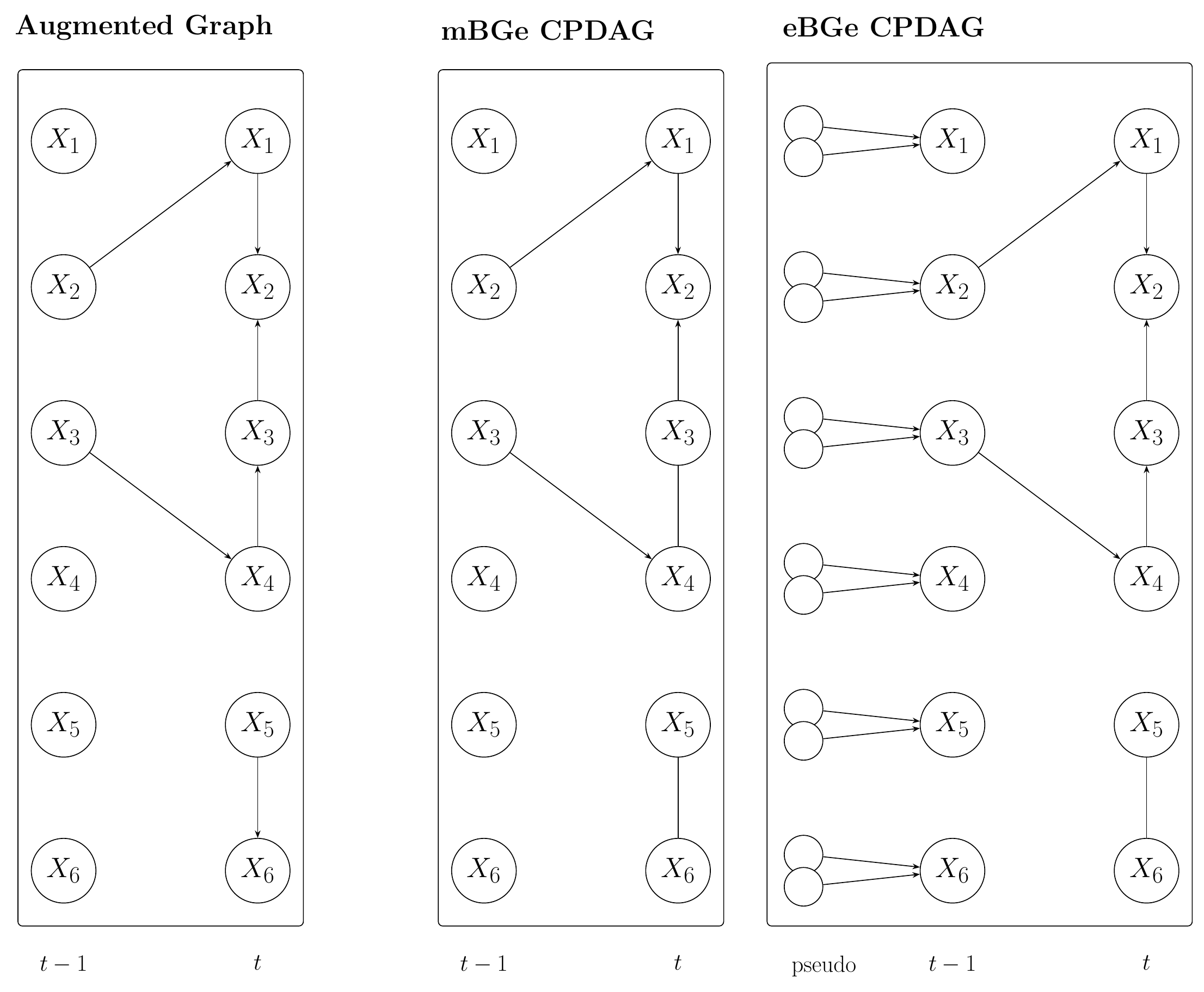}
\vspace{-0.2cm}
\caption{\footnotesize \textbf{Exemplary graphical illustration of the mBGe and eBGe CPDAGs.}
The left panel shows the augmented graph of a GDBN with $n=6$ nodes. The augmented graph contains
two dynamic edges and four static edges. The dynamic edges must point forward in time and are
therefore compelled. The CPDAG corresponding to the mBGe score is shown in the center panel; it
is obtained by merging the dynamic graph with the static CPDAG. The CPDAG of the enlarged graph (with pseudo parent nodes) is shown in the right panel. Two pseudo parent nodes are added to each node $X_{j,t-1}$ in order
to enforce all dynamic edges to be compelled. Extracting the CPDAG of the enlarged graph and
subsequently removing the $12$ pseudo parent nodes yields the CPDAG corresponding to the eBGe
score, which contains five directed edges and one undirected edge.}
\label{FIG6} 
\end{figure}

\noindent\textbf{DAG-to-CPDAG algorithm for the eBGe model:} \\
To determine the CPDAG of a graph learned under the eBGe model, we adapt the concept of
transition-sequence (TS) equivalent networks introduced by Tian and Pearl \cite{TS_1}.
Two augmented graphs $\mathcal{G}_1$ and $\mathcal{G}_2$ are equivalent under the eBGe model if
and only if they share (i) the same skeleton, (ii) the same set of v-structures, and
(iii) identical static parent sets for all nodes that possess at least one dynamic parent. For practical implementation, we borrow an idea from \cite{TS_2}: For each dynamic parent node $X_{j,t-1}$, we introduce
two pseudo parent nodes $D_{j,1}$ and $D_{j,2}$. These pseudo nodes form a v-structure converging
on $X_{j,t-1}$, $D_{j,1} \rightarrow X_{j,t-1} \leftarrow D_{j,2}$. \\
Then all dynamic edges of the form $X_{j,t-1}\rightarrow X_{i,t}$ are enforced to
be compelled. Indeed, reversing any such edge would introduce additional v-structures
\[
D_{j,1}\rightarrow X_{j,t-1}\leftarrow X_{i,t}
\quad\text{and}\quad
D_{j,2}\rightarrow X_{j,t-1}\leftarrow X_{i,t},
\]
which are not present in the original graph and therefore violate equivalence.
The resulting enlarged graph consists of the $2n$ nodes of the augmented graph together with
$2n$ additional pseudo parent nodes. Applying the standard DAG-to-CPDAG algorithm of
\cite{CHICK_2002} to this enlarged graph yields its CPDAG. Finally, removing the $2n$ pseudo
parent nodes produces the CPDAG corresponding to the eBGe model. A graphical illustration of this procedure is provided in Figure~\ref{FIG6}. \\

\noindent\textbf{Structural Hamming distances (eBGe vs.\ mBGe).} \\
We conduct a small simulation study to quantify the average difference between CPDAGs obtained
under the eBGe and mBGe scores. As a starting point, we consider the RAF protein signalling
pathway \cite{Sachs,WerhliBioinf}, which consists of $n=11$ nodes and $20$ edges. For different
values of $x$, we randomly select $x$ edges and declare them to be static, while the remaining
$20-x$ edges are declared to be dynamic. For each configuration, we extract the corresponding
mBGe and eBGe CPDAGs and compute their structural Hamming distance (SHD).\footnote{The SHD is the
number of edge insertions, deletions, and changes of orientation (including
directed--undirected substitutions) required to transform one CPDAG into another.}\\

Figure~\ref{FIG7} shows the average SHD for $x=0,\ldots,20$, where each average is taken over
$25$ replicates. The average SHD increases with $x$ and reaches a plateau in the second half of
the range. Two structural characteristics of the RAF pathway are that the maximum parent set
size is three and that all nodes are mutually connected by paths. To assess whether this behavior is specific to the RAF pathway, we also consider random DAG
structures with $n=11$ nodes and $20$ edges, of which $x$ randomly chosen edges are declared to
be static. The corresponding results are shown in Figure~\ref{FIG8}. In this case, the SHD
peaks at approximately $6$--$8$ static edges and exhibits a noticeably smoother profile.

\afterpage{

\begin{figure}[p]\centering
\includegraphics[width=0.99\linewidth]{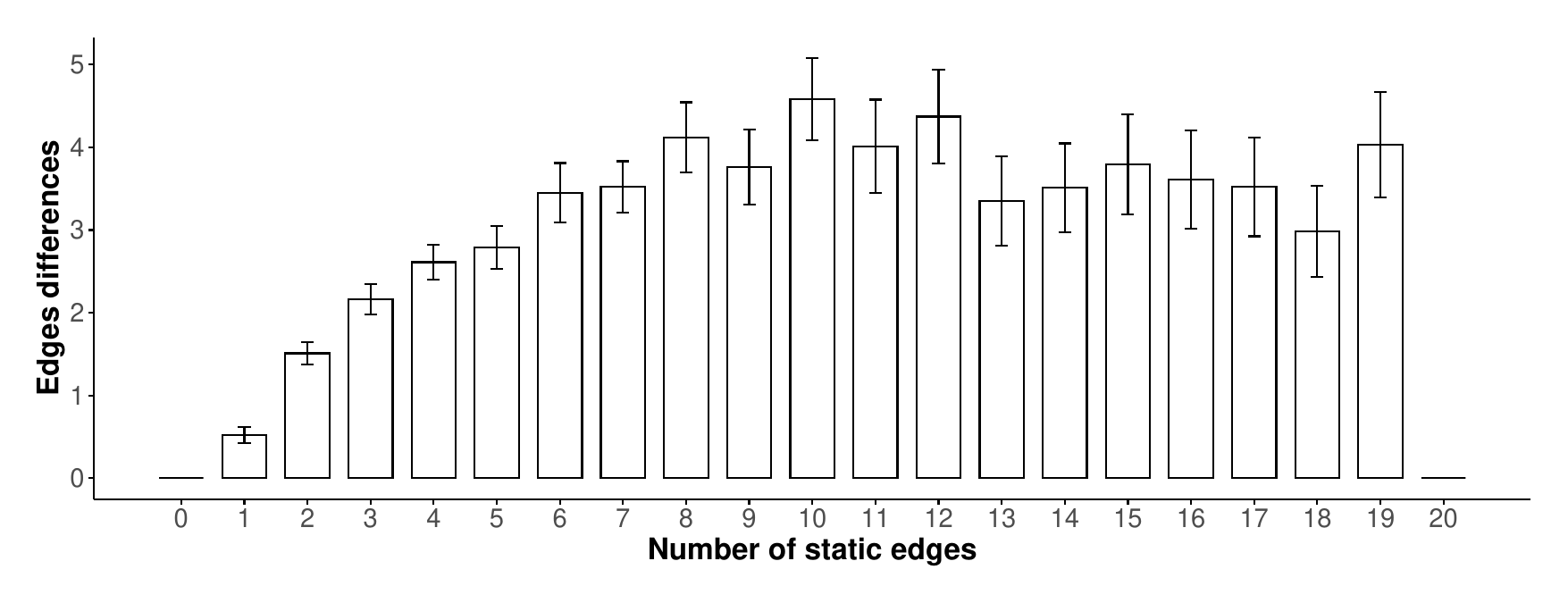}
\vspace{-0.5cm}
\caption{\footnotesize \textbf{Average structural Hamming distance (SHD) between the mBGe and eBGe CPDAGs for the RAF pathway with $\mathbf{n=11}$ nodes and $\mathbf{20}$ edges} \cite{Sachs,WerhliBioinf}. For each $x=0,1,\ldots,20$, we randomly selected $x$ edges and declared them to be static, while the remaining $20-x$ edges were declared to be dynamic. The SHDs are averaged over $25$ replicates, and the error bars indicate standard deviations.}
\label{FIG7}
\end{figure}

\begin{figure}[p]\centering
\includegraphics[width=0.99\linewidth]{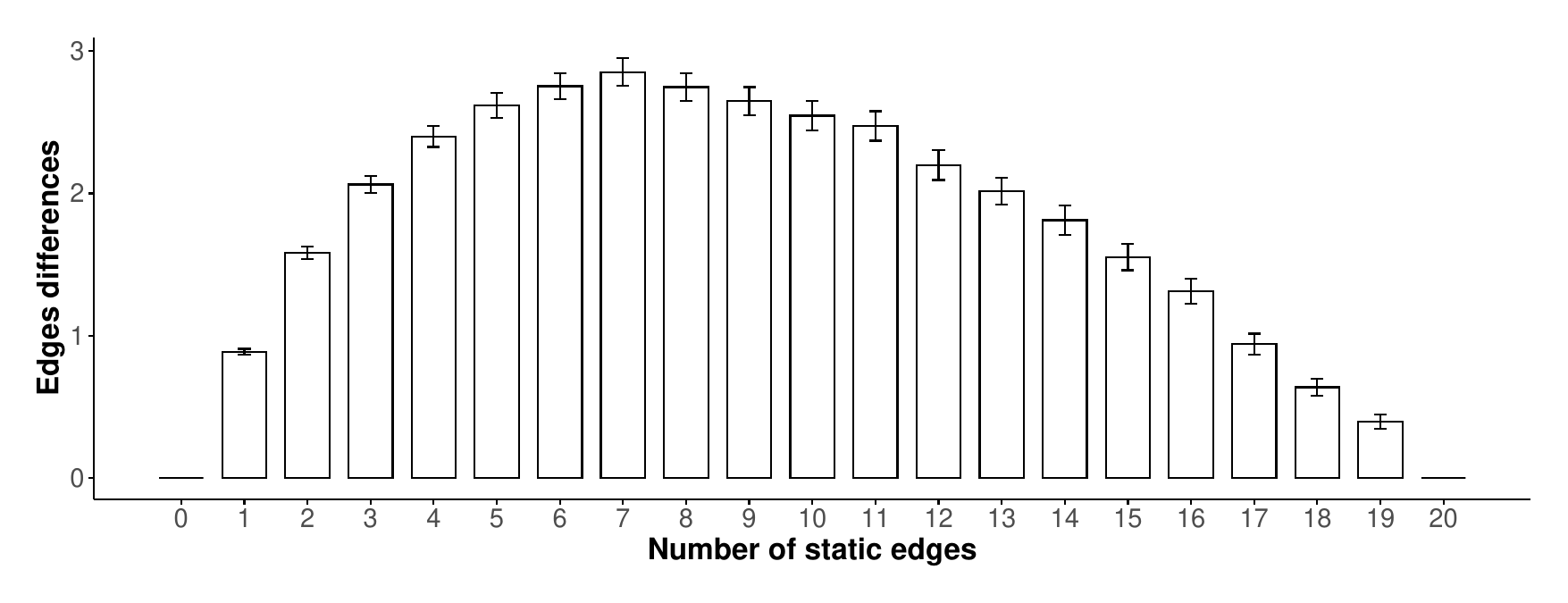}
\vspace{-0.5cm}
\caption{\footnotesize \textbf{Average structural Hamming distance (SHD) between the mBGe and eBGe CPDAGs for random DAGs with $\mathbf{n=11}$ nodes and $\mathbf{20}$ edges.} For each $x=0,1,\ldots,20$, we randomly selected $x$ edges and declared them to be static, while the remaining $20-x$ edges were declared to be dynamic. The SHDs are averaged over $25$ replicates, and the error bars indicate standard deviations.}
\label{FIG8}
\end{figure}
}

\section{Data}
\label{sec:data}
\subsection{Simulated data}
\label{section:simulated}

For the simulation study, we generated data under both the mBGe and eBGe models using a range of
network structures and tuning parameters, including network size, number of edges, regression
coefficient magnitudes, and noise variances. In this paper, we report empirical results for one
representative setting, in which we considered random DAGs with $n=11$ nodes and $20$ edges.
Among these edges, we declared $x\in\{5,10,15\}$ to be static, with the remaining $20-x$ edges
declared to be dynamic. Without loss of generality, we assume a topological ordering $X_1,\ldots,X_n$ of the DAG, so
that all edges are of the form $X_i \rightarrow X_j$ with $i<j$. For each node $X_i$, the DAG
determines a static parent set $\BPi_i^S$ and a dynamic parent set $\BPi_i^D$.

\subsubsection{Data of eBGe type}

For data generated under the eBGe model, observations are generated in a single step conditional on the previous time point. For
$t\geq 2$, we generate
\begin{equation}
\label{eq:ebge_data_gen}
X_{i,t}
=
\mu_{i,S}
+
\sum_{X_j \in \BPi_i^S}
\beta_{i,j}^S \bigl(X_{j,t}-\mu_{j,S}\bigr)
+
\sum_{X_k \in \BPi_i^D}
\beta_{i,k}^D \bigl(X_{k,t-1}-\mu_{k,D}\bigr)
+
\epsilon_{i,t},
\end{equation}
where $\beta_{i,j}^S$ and $\beta_{i,k}^D$ denote the regression coefficients associated with
static and dynamic edges, respectively, and $\mu_{i,S}$ and  $\mu_{i,D}$  are marginal mean parameters. The noise
terms $\epsilon_{i,t}$ are independent and identically distributed. For $t=2,\ldots,T$, the values $X_{1,t},\ldots,X_{n,t}$ are generated sequentially, respecting
both the temporal ordering and the topological ordering of the DAG.

\subsubsection{Data of mBGe type}

For data generated under the mBGe model, observations are generated in two successive steps.
First, we generate static Bayesian network data for the mean-adjusted variables
\[
Y_{i,t} := X_{i,t} - \mu_{i,t}^{\star},
\]
and subsequently generate and add the time-varying means $\mu_{i,t}^{\star}$. For the mean-adjusted variables, we have for each $t\geq 1$
\begin{equation}
\label{mean_adjusted}
Y_{i,t}
=
\sum_{X_j \in \BPi_i^S} \beta_{i,j}^S\, Y_{j,t}
+
\epsilon_{i,t},
\end{equation}
where $\beta_{i,j}^S$ denote the regression coefficients associated with the static parent set
$\BPi_i^S$. For each time point $t\geq 1$, the values $Y_{1,t},\ldots,Y_{n,t}$ are generated
sequentially in topological order. \\

In a second step, we generate the time-varying mean components. For each $t\geq 2$, the mean of
$X_i$ is given by
\begin{equation}
\label{eq:mbge_mean}
\mu_{i,t}^{\star}
=
\beta_{i,0}^D
+
\sum_{X_k \in \BPi_i^D} \beta_{i,k}^D\, Y_{k,t-1},
\end{equation}
where $\beta_{i,k}^D$ denote the regression coefficients associated with the dynamic parent set
$\BPi_i^D$. Combining the two steps yields
\begin{equation}
\label{eq:mbge_data_gen}
X_{i,t}
=
\beta_{i,0}^D
+
\sum_{X_j \in \BPi_i^S} \beta_{i,j}^S \bigl(X_{j,t}-\mu_{j,t}^{\star}\bigr)
+
\sum_{X_k \in \BPi_i^D} \beta_{i,k}^D \bigl(X_{k,t-1}-\mu_{k,t-1}^{\star}\bigr)
+
\epsilon_{i,t}.
\end{equation}

This expression highlights the key difference between data generated under the mBGe and eBGe
models. For mBGe data, the mean parameters $\mu_{j,t}^{\star}$ and $\mu_{k,t-1}^{\star}$ are
time-varying. In particular, the mean of $X_j$ at time $t$, denoted by $\mu_{j,t}^{\star}$,
depends on the values of the dynamic parents of $X_j$ at the previous time point $t-1$.

\subsubsection{Parameter settings}
The absolute values of all regression coefficients $\beta_{i,j}^S$ and $\beta_{i,j}^D$ were
sampled independently from a uniform distribution on the interval $[0.5,2]$, and random signs
were assigned to each coefficient. All intercept terms ($\mu_{i,S}$, $\mu_{i,D}$, and $\beta_{i,0}^D$) were set to zero. The error terms were assumed to be Gaussian with zero mean and variance $4$,
that is, $\epsilon_{i,t}\sim\mathcal{N}(0,4)$. For the initial time point $t=1$, dynamic parent effects were ignored. Specifically, for the
eBGe model we set $X_{i,0}=0$ for all $i$ as an initial condition, and for the mBGe model we set
$\mu_{i,1}^{\star}=0$ for all $i$, so that $X_{i,1}=Y_{i,1}$.

\subsection{Real-world datasets}
\label{section:real}

We compared the performance of the mBGe and eBGe scores on five real-world datasets. For a time
series observed at time points $t=1,\ldots,T$, the dynamic parent values required at $t=1$ are
unavailable, so that the effective sample size is reduced to $T^{\star}=T-1$. If a dataset
consists of measurements from multiple independent experiments $k=1,\ldots,K$, where experiment
$k$ spans $T_k$ time points, then each experiment contributes $T_k-1$ usable observations. These
observations can be concatenated to form a single time series with total sample size
$T^{\star}=\sum_{k=1}^{K}(T_k-1)$. The five real-world datasets are as follows.

\medskip
\noindent\textbf{mTOR protein signalling data.}
The mammalian target of rapamycin (mTOR) protein signalling pathway plays a key role in the
pathophysiology of tumours. The dataset from \cite{KATHRIN} contains measurements of $n=11$
phosphorylation sites across eight proteins, collected in $K=2$ independent experiments, each
consisting of $T_k=10$ time points.

\medskip
\noindent\textbf{\textit{Arabidopsis thaliana} gene expression data.}
The \textit{Arabidopsis thaliana} gene expression data from \cite{GrzHusNIPS2009} contain
measurements of $n=9$ genes involved in the circadian clock. The genes were measured in
$K=4$ independent experiments with $T_k\in\{12,13\}$.

\medskip
\noindent\textbf{Yeast gene expression data.}
The yeast gene expression data from \cite{CantoneBernadoCell} contain measurements of $n=5$ genes
involved in galactose and glucose metabolism. The genes were measured in $K=2$ independent
experiments with $T_k\in\{16,21\}$ time points.

\medskip
\noindent\textbf{Andromeda data.}
The Andromeda dataset from \cite{HATZIKOS} contains measurements of $n=6$ water-quality variables
collected in the Thermaikos Gulf of Thessaloniki, Greece. The variables were measured in a single
experiment ($K=1$) at $T=59$ time points.

\medskip
\noindent\textbf{Occupational Employment Survey 2010 (OES10).}
This final application differs from the previous four examples in that the records of the
Occupational Employment Survey 2010 (OES10) are not time-ordered. The OES10 data contain counts of
full-time equivalent employees across different job categories and U.S.\ cities. Following
\cite{SALAM_CS}, we treat cities as observations and job categories as variables, and use the
data to learn Gaussian Bayesian networks with additional external variables. Specifically, we
learn Gaussian Bayesian networks for $n=10$ variables $X_{1,t},\ldots,X_{n,t}$, where the index
$t$ refers to different cities. Since there is no temporal structure, no lagged variables
$X_{1,t-1},\ldots,X_{n,t-1}$ are available as dynamic parent nodes. Instead, we select ten
additional variables $Z_{1,t},\ldots,Z_{n,t}$ and treat them as potential external parent nodes.

\section{Implementation details}
\label{sec:technical}

For the simulation studies in Section~\ref{sec:results}, we employ uniform graph priors,
$P(\mathcal{G})=c_1$ and $P(\mathcal{G}^D)=c_2$ for all $\mathcal{G}$ and $\mathcal{G}^D$, and we rule out self-loops of the form $X_{i,t-1}\rightarrow X_{i,t}$. For the
parameter priors, we choose weakly informative hyperparameters. 
For the mBGe model, we use the $n$-dimensional Wishart prior from
Eq.~(\ref{BGE_prior_mix_old}) and set $\alpha_w=n+2$ and ${\bf R}={\bf I}$. For the regression
coefficients, we use a standard Gaussian prior by setting $\lambda^2=1$ in
Eq.~(\ref{eq_reg_prior}). For the eBGe model, we use the $2n$-dimensional Normal--Wishart prior from
Eqs.~(\ref{eq_aug_p1}--\ref{eq_aug_p2}) and set $\alpha_w=2n+2$, $\tilde{\mathbf{R}}=\mathbf{I}$,
$\tilde{\boldsymbol{\nu}}=\mathbf{0}$, and $\alpha_{\mu}=1$. \\
To ensure the appropriateness of these hyperparameter choices, all variables were
$z$-score standardized to have mean zero and variance one.

\medskip
\noindent\textbf{Simulation lengths and computational costs.}\\
We ran all MCMC simulations for $100{,}000$ iterations. Discarding the first $50\%$ of samples as
burn-in and thinning the remainder by a factor of $100$ yielded posterior samples of size $500$.
For networks of moderate size, these settings resulted in satisfactory convergence. Convergence
was assessed using standard diagnostics, including trace plots and scatter plots of edge
inclusion probabilities. The average computational costs for $100{,}000$ MCMC iterations of the eBGe and mBGe models are
reported in Table~\ref{tab:comp_costs} in Appendix~\ref{sec:appendix}. In contrast to the eBGe
model, whose computational cost remains consistently low across all settings, the cost of the
mBGe model is substantially higher and increases approximately exponentially with the sample
size~$T$, as reported in Table~\ref{tab:comp_costs} in \ref{sec:appendix}.

\medskip
\noindent\textbf{Edge inclusion probabilities and predictive probabilities.}\\
For the simulated data in Section~\ref{section:simulated}, our primary focus was on recovering
the underlying network structure. To this end, each posterior-sampled DAG was first converted
into its corresponding mBGe or eBGe CPDAG. From the resulting CPDAG sequence, we computed
marginal edge inclusion probabilities. Specifically, for each possible edge $X_i\rightarrow X_j$,
we computed the proportion of posterior CPDAGs in which this edge appeared. Since undirected CPDAG edges $X_i-X_j$ indicate uncertainty about the edge direction, we treated
such edges as bidirectional ($X_i\leftrightarrow X_j$) when computing inclusion probabilities.\\
For the real-world datasets in Section~\ref{section:real}, we used posterior samples of the
network parameters to compute Bayesian predictive probabilities for held-out observations within
a leave-one-out cross-validation (LOOCV) framework. Specifically, for each held-out observation,
we ran an MCMC simulation on the remaining data, obtained posterior draws of the network
parameters, and used these draws to compute the predictive probability of the held-out
observation.

\begin{figure}[p]\centering
\includegraphics[width=0.99\linewidth]{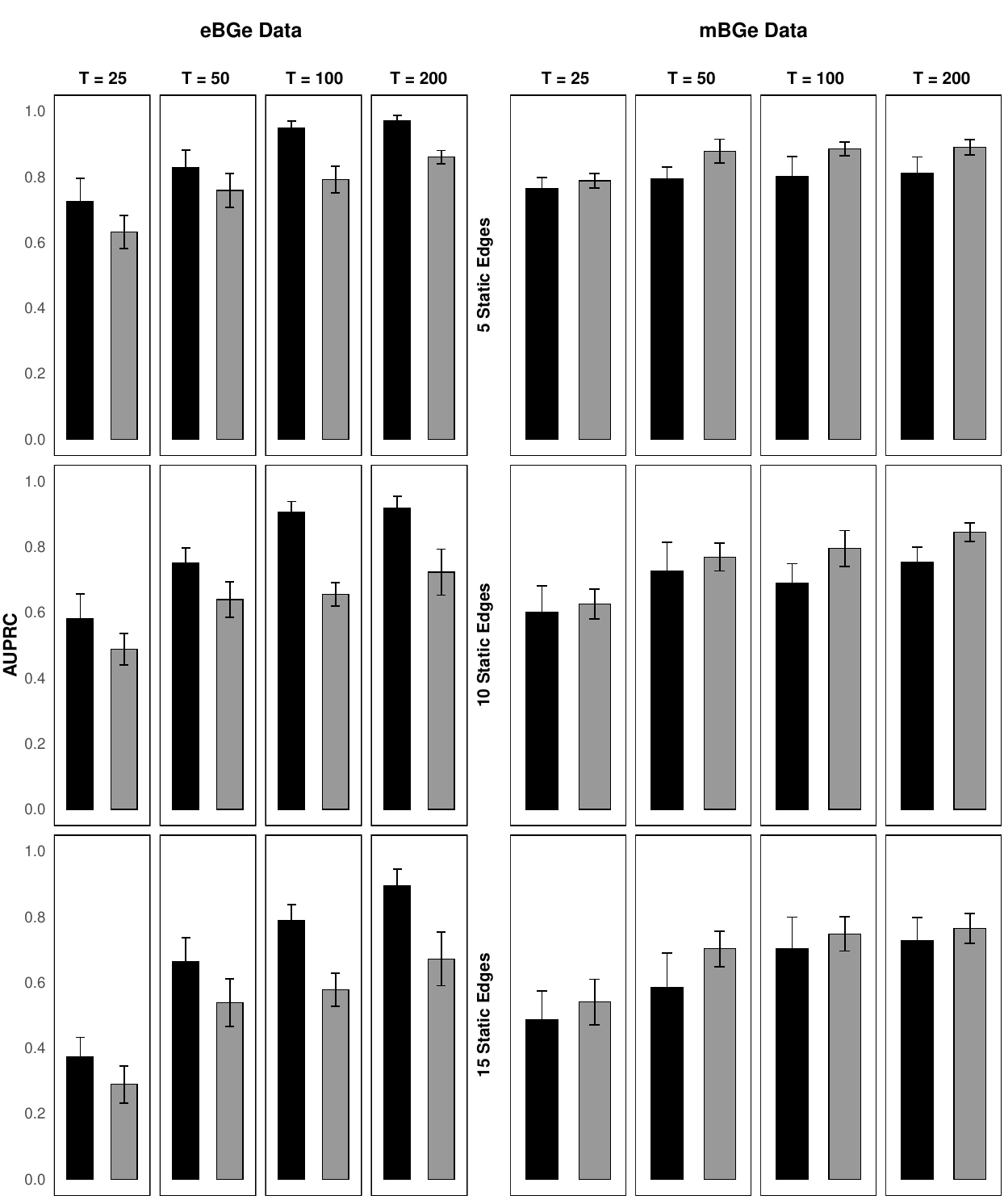}
\vspace{-0.25cm}
\caption{\footnotesize \textbf{Average area under the precision--recall curve (AUPRC) for simulated network data.} Each AUPRC value is averaged over 10 independent data realizations, with error bars indicating 95\% confidence intervals. Bars show the AUPRC achieved by the eBGe (black) and mBGe (grey) models. Data were simulated from random DAGs with $n=11$ nodes and 20 edges, using either the eBGe model (left panel) or the mBGe model (right panel). The time-series length was varied over $T\in\{25,50,100,200\}$, and $x\in\{5,10,15\}$ of the 20 edges were randomly selected to be static, with the remaining $20-x$ edges being dynamic. For further details, see Section~\ref{section:simulated}.}
\label{FIG9}
\end{figure}

\afterpage{

\begin{figure}[p]\centering
\includegraphics[width=0.99\linewidth]{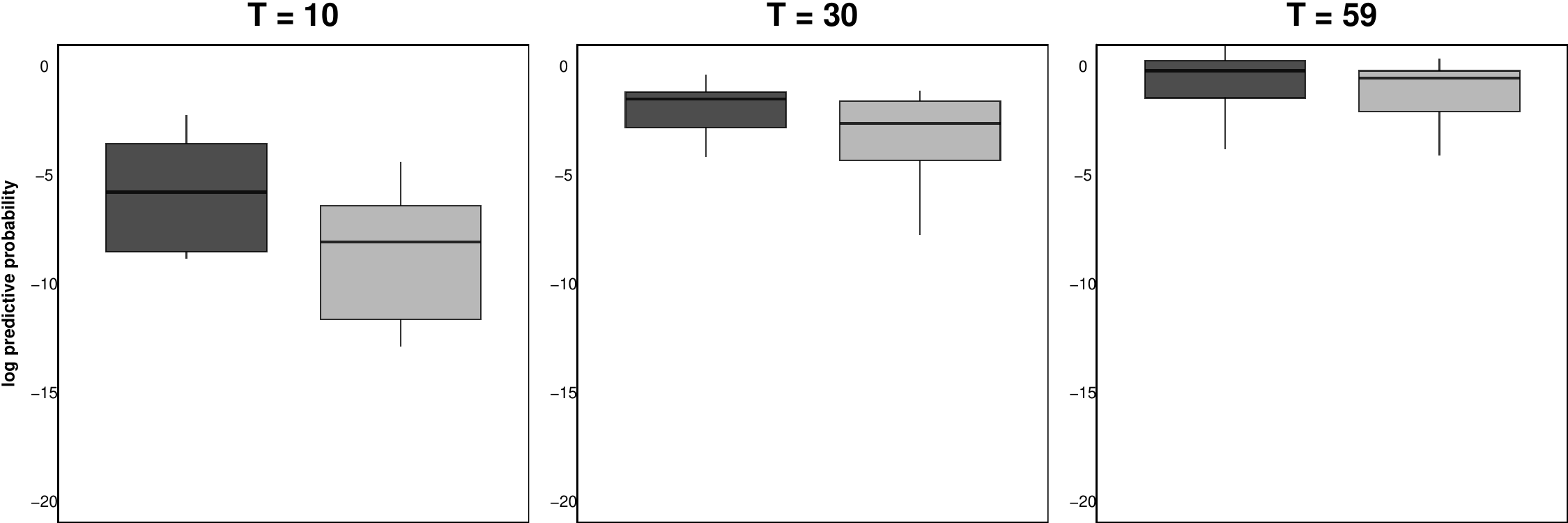}
\vspace{-0.3cm}
\caption{\footnotesize \textbf{Predictive probabilities for the Andromeda data.} Boxplots show the predictive probabilities obtained with the eBGe (dark) and mBGe (light grey) models. The sample size $T$ was varied, and predictive probabilities were computed using LOOCV.}
\label{FIG10}
\end{figure}

\begin{figure}[p]\centering
\includegraphics[width=0.99\linewidth]{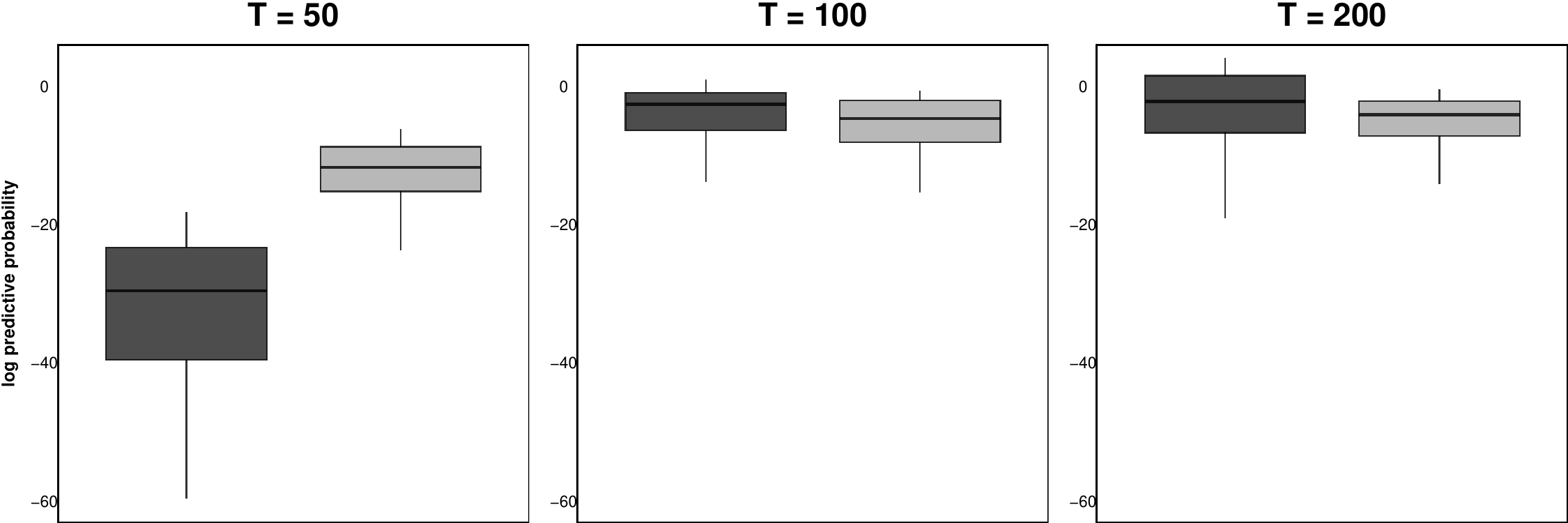}
\vspace{-0.25cm}
\caption{\footnotesize \textbf{Predictive probabilities for the OES10 data.} Boxplots show the predictive probabilities obtained with the eBGe (dark) and mBGe (light grey) models. The sample size $T$ was varied, and predictive probabilities were computed using LOOCV.}
\label{FIG11}
\end{figure}

\begin{figure}[p]\centering
\includegraphics[width=0.99\linewidth]{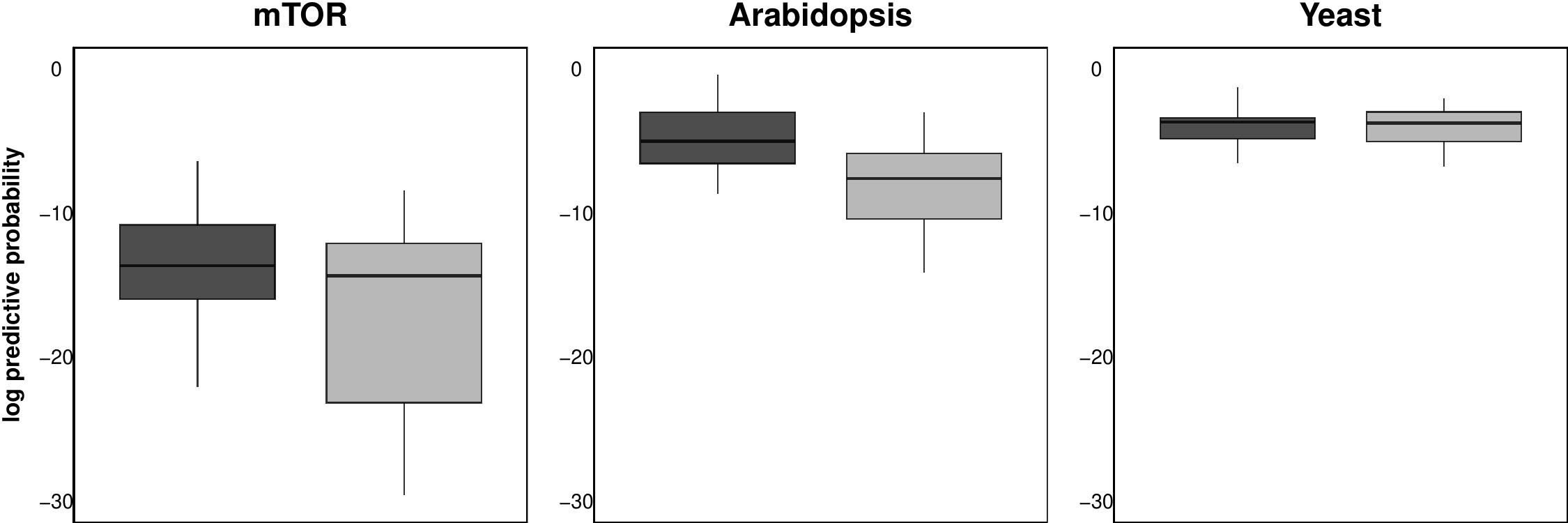}
\vspace{-0.25cm}
\caption{\footnotesize \textbf{Predictive probabilities for the mTOR, Arabidopsis, and Yeast data.} Boxplots show the log predictive probabilities obtained with the eBGe (dark) and mBGe (light grey) models. Predictive probabilities were computed using LOOCV.}
\label{FIG12}
\end{figure}

}

\section{Empirical Results}
\label{sec:results}
We first apply both methods (eBGe and mBGe) to simulated data in order to compare their network
reconstruction accuracies. To this end, we generate data from a range of random network
structures with $n=11$ nodes, $20$ edges, and randomly sampled network parameters. A detailed
description of the data generation procedure is provided in
Section~\ref{section:simulated}. For each of the 12 combinations of three numbers of static edges
$x\in\{5,10,15\}$ and four time-series lengths $T\in\{25,50,100,200\}$, we generate ten datasets
from the mBGe model and ten datasets from the eBGe model, resulting in a total of 240 datasets.
Figure~\ref{FIG9} reports the average network reconstruction accuracies in terms of the area under
the precision--recall curve (AUPRC).\footnote{AUPRC scores are obtained by computing posterior edge
inclusion probabilities, ranking edges according to these probabilities, constructing the
corresponding precision--recall curve using the known ground-truth networks, and computing the
area under this curve.} The AUPRC results align with our expectations: the mBGe model performs
consistently worse when applied to data generated under the eBGe model, and the eBGe model
performs consistently worse when applied to data generated under the mBGe model. Interestingly,
applying the mBGe model to eBGe data leads to a larger performance degradation than applying the
eBGe model to mBGe data.\\
For the five real-world datasets introduced in Section~\ref{section:real}, we compare the two
models in terms of Bayesian predictive probabilities.\footnote{Bayesian predictive probabilities
are computed as posterior predictive densities by integrating the likelihood of the held-out
observation over the posterior distribution of the network parameters, approximated via Monte
Carlo averaging over posterior samples.} We use Bayesian predictive probabilities because the
true underlying network structures are unknown for real-world data. A leave-one-out
cross-validation scheme is employed, where each observation is held out once and the remaining
data are analysed to compute the predictive probability of the held-out observation. More
specifically, network parameters are sampled from the posterior distribution based on the
reduced dataset, and the predictive probability is obtained by averaging the likelihood of the
held-out observation over these samples. The results for the Andromeda and OES10 datasets are shown in
Figures~\ref{FIG10} and~\ref{FIG11}. In both applications, we vary the number of observations.
For the Andromeda data, we consider the initial $T\in\{15,30,59\}$ out of $T=59$ time points.
For the OES10 data, we select $T\in\{50,100,200\}$ cities. Figure~\ref{FIG12} shows boxplots of
the Bayesian predictive probabilities for the mTOR, Arabidopsis, and Yeast datasets.
Overall, no clear trend emerges, although the eBGe model tends to perform slightly better in
terms of Bayesian predictive probabilities. Across the nine dataset–sample-size combinations considered, eBGe performs worse than mBGe in only one case (OES10 with $T=50$). By contrast, eBGe performs better for all three sample sizes of the Andromeda data as well as for the Arabidopsis data.

\section{Discussion and conclusions}
\label{sec:conclusions}
We compared two Bayesian approaches to Gaussian dynamic Bayesian networks (GDBNs)
with inter- and intra-slice edges. We showed that the two models, referred to as the extended BGe
(eBGe) and the mean-adjusted BGe (mBGe), induce different graph equivalence classes. While
the equivalence classes implied by the mBGe model can be obtained using standard DAG-to-CPDAG
algorithms \cite{Chickering_UAI95,CHICK_2002}, this is not the case for the eBGe model. To address
this issue, we demonstrated how existing DAG-to-CPDAG algorithms can be adapted in a
straightforward manner to correctly identify equivalence classes under the eBGe model. The
proposed modification is also required for identifying equivalence classes of discrete dynamic
Bayesian networks. Since the required changes are conceptually simple, they can be incorporated
into existing Bayesian network software packages such as \emph{bnlearn} and \emph{BiDAG} with
minimal effort. Although the eBGe model may be subject to a small bias, as certain parameter constraints cannot be fully accounted for (cf.\ Section~\ref{sec:ebge}), our empirical results in
Section~\ref{sec:results} suggest that the eBGe model generally exhibits superior performance.
Moreover, inference under the eBGe model is substantially more computationally efficient. By contrast, the computational cost of the mBGe model appears to increase approximately exponentially with the time-series length~$T$. \\
 

\noindent {\bf Acknowledgments:} \\ Kezhuo Li acknowledges the support of the China Scholarship Council (Project
ID: 202306310059).
\newpage
\appendix

\section{Computational Costs}
\label{sec:appendix}
For the simulated network data described in Section~\ref{section:simulated}, we also recorded the computational costs in terms of wall-clock time. The computational costs for 100,000 (100k) MCMC iterations under the eBGe and mBGe models are reported in Table~\ref{tab:comp_costs}. All simulations were carried out using our own {\bf R} code on a computing cluster equipped with Intel Xeon 2.5 GHz nodes and 4 GB of RAM. We emphasize that considerably more efficient implementations are available. For example, a substantially faster implementation of the eBGe model is available in the {\bf R} package {\em BiDAG} \citep{SOFTWARE3}. As can be seen from Table~\ref{tab:comp_costs}, the computational costs for the eBGe model remain consistently low across all settings, whereas the costs for the mBGe model are substantially higher and appear to increase roughly exponentially as a function of $T$.


\begin{table}[h]
\centering
\begin{tabular}{ccccc}
\toprule
\multirow{2}{*}{$x$} & \multicolumn{2}{c}{$T = 25$} & \multicolumn{2}{c}{$T = 50$} \\
\cmidrule(lr){2-3} \cmidrule(lr){4-5}
 & eBGe & mBGe & eBGe & mBGe \\
\midrule
5 & 1 m 23 s & 10 m 40 s & 1 m 25 s & 18 m 31 s \\
 & $(\pm 0.82\text{ s})$ & $(\pm 4.11\text{ s})$ & $(\pm 0.75\text{ s})$ & $(\pm 3.65\text{ s})$ \\[0.3em]
10 & 1 m 24 s & 11 m 18 s & 1 m 26 s & 18 m 40 s \\
 & $(\pm 0.69\text{ s})$ & $(\pm 2.86\text{ s})$ & $(\pm 0.87\text{ s})$ & $(\pm 4.89\text{ s})$ \\[0.3em]
15 & 1 m 28 s & 11 m 46 s & 1 m 29 s & 19 m 18 s \\
 & $(\pm 1.08\text{ s})$ & $(\pm 3.38\text{ s})$ & $(\pm 0.46\text{ s})$ & $(\pm 7.49\text{ s})$ \\
\midrule
\multirow{2}{*}{$x$} & \multicolumn{2}{c}{$T = 100$} & \multicolumn{2}{c}{$T = 200$} \\
\cmidrule(lr){2-3} \cmidrule(lr){4-5}
 & eBGe & mBGe & eBGe & mBGe \\
\midrule
5 & 1 m 27 s & 24 m 01 s & 1 m 28 s & 63 m 21 s \\
 & $(\pm 0.58\text{ s})$ & $(\pm 10.01\text{ s})$ & $(\pm 0.47\text{ s})$ & $(\pm 11.61\text{ s})$ \\[0.3em]
10 & 1 m 28 s & 25 m 58 s & 1 m 29 s & 64 m 28 s \\
 & $(\pm 0.39\text{ s})$ & $(\pm 5.19\text{ s})$ & $(\pm 0.55\text{ s})$ & $(\pm 13.10\text{ s})$ \\[0.3em]
15 & 1 m 31 s & 26 m 15 s & 1 m 39 s & 64 m 45 s \\
 & $(\pm 0.73\text{ s})$ & $(\pm 12.07\text{ s})$ & $(\pm 1.28\text{ s})$ & $(\pm 14.55\text{ s})$ \\
\bottomrule
\end{tabular}
\caption{\label{tab:comp_costs} \footnotesize {\bf Computational costs for MCMC simulations for the eBGe and mBGe models:} Reported are minutes [m] and seconds [s], together with standard deviations $(\pm\text{sd})$, for 100k MCMC iterations.  The table distinguishes twelve combinations of the number of static edges ($x \in \{5,10,15\}$) and the number of observations ($T \in \{25,50,100,200\}$).  All values are averaged over ten MCMC runs on different data instantiations.}
\end{table}

\bibliographystyle{elsarticle-num-names} 
\bibliography{tdh}

\end{document}